\documentclass{book}
\usepackage{epsfig}

\newcommand{\code}[1]{\texttt{#1}}

\newcounter{figctr}

\newcounter{tabctr}

\usepackage[cip,ChapterTOCs]{sunil1}
\usepackage{floatflt}
\usepackage{amssymb}
\usepackage{amsmath}
\usepackage{latexsym}
\usepackage{theorem}
\usepackage{graphicx}
\usepackage{subfigure}
\usepackage{graphics}
\usepackage{epsfig}
\usepackage{url}
\usepackage{makeidx}
\usepackage{showidx}

\usepackage{multicol}

\makeatletter
\newbox\tempbox
\newdimen\nomenwidth

\def\symbolentry#1#2{\par\noindent\@hangfrom{\hbox to \nomenwidth{#1\hss}}#2\par}
\makeatother

\newcommand\ez{\texttt{Enzo}}

\begin{document}
\tableofcontents


\title{Petascale Computing: Algorithms and Applications}

\chapterauthor{Michael L. Norman, James Bordner, Daniel Reynolds and Rick Wagner}{Laboratory for Computational Astrophysics, Center for Astrophysics and Space Sciences, University of California San Diego, 9500 Gilman Dr., La Jolla CA 92093, (mlnorman, jobordner, drreynolds, rpwagner)@ucsd.edu}
\chapterauthor{Greg L. Bryan}{Astronomy Department, Columbia University, 550 West 120th St., New York NY 10027, gbryan@astro.columbia.edu}
\chapterauthor{Robert Harkness}{San Diego Supercomputer Center, 9500 Gilman Dr., La Jolla CA 92093, harkness@sdsc.edu}
\chapterauthor{Brian O'Shea}{Theoretical Astrophysics (T-6), Los Alamos National Laboratory, P.O. Box 1663, Los Alamos, NM 87545, bwoshea@lanl.gov}

\chapter{Simulating Cosmological Evolution with Enzo}

\newcommand{\pargraph}[1]{\devel{\P\ \textbf{#1} \\}}
\newcommand{\devel}[1]{#1}  
\section{Cosmological structure formation}
The universe is homogeneous and isotropic on scales exceeding half a
billion light years, but on smaller scales it is clumpy, exhibiting a
hierarchy of structures ranging from individual galaxies up to groups and
clusters of galaxies, and on the largest scales, the galaxies are
aligned in a cosmic web of filaments and voids.
In between the galaxies is a diffuse
plasma which is the reservoir of matter out of which galaxies form.
Understanding the origin and evolution of these structures is the goal of {\em
cosmological structure formation} (CSF). 

It is now understood that CSF
is driven by the gravitational clustering of dark matter, the dominant
mass constituent of the universe. Slight inhomogeneities in the dark matter
distribution laid down in the early universe seed CSF. The rate at which
structure develops in the universe depends upon the power spectrum of
these perturbations, now well measured on scales greater than one million light years \cite{Spergel06}.
It also depends on the cosmic expansion rate,
which in turn is influenced by dark energy, a form of energy which
exhibits negative pressure and whose existence was
revealed only in 1998 by the discovery that the expansion rate is
accelerating. The quantitative study of CSF is thus a direct route to
studying two of the most mysterious substances in modern physics: dark
matter and dark energy.

The part of the universe that astronomers can see directly is made up of
ordinary ``baryonic" matter. Thus, in order to make contact with
observations, we must simulate the detailed dynamics and
thermodynamics of the cosmic plasma---mostly hydrogen and helium---under
the influence of dark matter and dark energy, as well as ionizing
radiation backgrounds. Such simulations are called hydrodynamic
cosmological simulations, and are what we consider in this chapter. CSF is
inherently nonlinear, multidimensional, and involves a variety of
physical processes operating on a range of length- and
time-scales. Large scale numerical simulation is the primary means we
have of studying it in detail.

\begin{figure}
\begin{center}
\includegraphics[width=\textwidth]{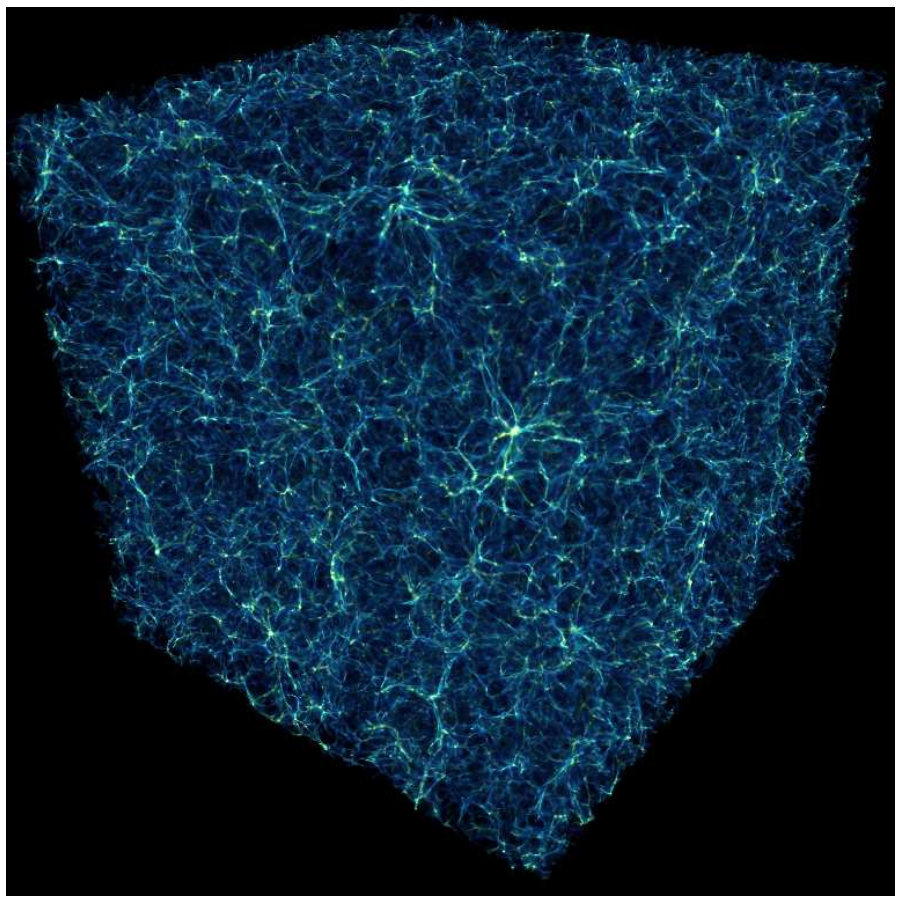}
\end{center}
\caption{\ez\ hydrodynamic simulation of cosmic structure in a 700 Mpc volume
of the universe. Up to seven levels of adaptive mesh refinement resolve the distribution of baryons within and between galaxy clusters, for an effective resolution of
$65,536^3$. Volume rendering of baryon density. Image credit: M. Hall, NCSA}
\label{fig.cube}
\end{figure}

To give a feeling for the range of scales involved, the large scale distribution of galaxies in the present universe traces out a web-like pattern on typical scales
of 100 million light years. Individual galaxies are $10^{3-4}$ times smaller in
linear scale. Resolving the smallest galaxies with ten resolution elements per
radius yields a range of scales of $2 \times 10^5$ throughout the large scale volume.
A uniform grid of this size in 3D is out of the question, now and in the near future. However, such a high dynamic range is not needed everywhere, but only where the galaxies are located.
Using adaptive mesh refinement (AMR) techniques we are close to achieving this dynamic range  running on today's terascale computers including a simple decription of the baryonic fluid (Fig. \ref{fig.cube}). With petascale platforms, even
higher dynamic ranges will be achievable including the complex baryonic physics
that governs the formation and evolution of galaxies.  

In this paper we describe \ez \cite{enzo} a multiphysics, parallel, AMR application for simulating CSF developed at UCSD and Columbia. We describe its
physics, numerical algorithms, implementation, and performance on current terascale platforms. We also discuss our future plans
and some of the challenges we face as we move to the petascale.
%
%
\section{The Enzo code}
\subsection{Physical model and numerical algorithms}
Matter in the universe is of two basic types:
baryonic matter composed of atoms and molecules out of
which stars and galaxies are made, and non-baryonic ``dark'' matter of
unknown composition, which is nevertheless known to be the dominant
mass constituent in the universe on galaxy scales and
larger. \ez\ self-consistently simulates both components, which
evolve according to different physical laws and therefore require
different numerical algorithms. 

Baryonic matter is evolved using a finite volume discretization of the
Euler equations of gas dynamics cast in a frame which expands with the
universe. Energy source and sink terms due to radiative heating
and cooling processes are included, as well as changes in ionization state of the
gas \cite{azan97}. We use the Piecewise Parabolic Method (PPM), which
is a higher-order Godunov scheme developed by Colella and Woodward for
ideal gas dynamics calculations \cite{Bryan95}. The species abundances for H, H+, He,
He+, He++, and e- (and optionally H$_2$, HD and related species) 
are solved out of equilibrium by integrating the
rate equations including radiative and collisional
processes \cite{azan97}. Radiation fields are modeled as evolving but spatially
homogeneous backgrounds using published prescriptions.

Dark matter is assumed to behave as a collisionless phase fluid,
obeying the Vlasov-Poisson equation. Its evolution is solved using
particle-mesh algorithms for collisionless N-body dynamics
\cite{HE88}. In particular, we use the spatially second order-accurate
Cloud-in-Cell (CIC) formulation, together with leapfrog time
integration, which is formally second order-accurate in time. Dark
matter and baryonic matter interact only through their self-consistent
gravitational field. The gravitational potential is computed by
solving the Poisson equation on the uniform or adaptive grid hierarchy
using Fast Fourier Transform and multigrid techniques. In generic
terms, \ez\ is a 3D hybrid code consisting of a multi-species
hydrodynamic solver for the baryons coupled to a particle-mesh solver
for the dark matter via a Poisson solver.

Matter evolution is computed in a cubic domain of length $L=a(t)X$,
where $X$ is the domain size in comoving coordinates, and $a(t)$ is
the homogenous and isotropic scale factor of the universe which is an
analytic or numerical solution of the Friedmann equation, a first
order ODE. For sufficiently large $X$ compared to the structures of
interest, any chunk of the universe is statistically equivalent to any
other, justifying the use of periodic boundary conditions. The speed
of FFT algorithms and the fact that they are
ideally suited to periodic problems make them the Poisson solver of
choice given the large grids employed---$1024^3$ or larger.

CSF simulations require very large grids and particle numbers due to
two competing demands: large boxes are needed for a fair statistical
sample of the universe; and high mass and spatial resolutions are
needed to adequately resolve the scale lengths of the structures which
form. For example, in order to simultaneously describe galaxies' large scale
distribution in space (large scale structure) and adequately resolve their internal
structure, a dynamic range of
$10^5$ per spatial dimension and $10^9$ in mass is needed {\em at a
minimum}, as discussed above.

\subsection{Adaptive mesh refinement}

The need for higher resolution than afforded by uniform grids
motivated the development of \ez. \ez\ uses structured adaptive
mesh refinement (SAMR, \cite{Berger89,Bryan98}) to achieve high
resolution in gravitational condensations. The central idea behind
SAMR is simple to describe but challenging to implement efficiently on
parallel computers. The idea is this: while solving the desired set of equations on a
coarse uniform grid, monitor the quality of the solution and when
necessary, add an additional, finer mesh over the region that requires
enhanced resolution.  This finer (child) mesh obtains its boundary
conditions from the coarser (parent) grid or from other neighboring
(sibling) grids with the same mesh spacing. The finer grid is also
used to improve the solution on its parent.  In order to simplify the
bookkeeping, refined patches are required to have a cell spacing
which is an integer number divided by the parent's spacing.  In
addition, refined patches must begin and end on a parent cell
boundary.  As the evolution
continues, it may be necessary to move, resize or even remove the
finer mesh.  Refined patches themselves may require further refinement,
producing a tree structure that can continue to any depth.
We denote the level of a patch by the number of times it has been
refined compared to the root grid.  If the cell spacing of the root grid
(level 0) is $\Delta x$, then the cell spacing of a mesh at level $l$ is
$\Delta x / r^{l}$ where $r$ is the integer refinement factor (typically 2).

To advance our system of coupled equations in time on this grid
hierarchy, we use a recursive algorithm. The {\tt EvolveLevel} routine
is passed the level of the hierarchy it is to work on and the new
time. Its job is to march the grids on that level from the old time to
the new time:

\begin{figure}
\begin{minipage}{\textwidth}
\begin{tabbing}
xxxx\=xxxx\=xxxx\= \kill
\> \code{EvolveLevel(level, ParentTime)} \\
\> \code{begin} \\
\>\>   \code{SetBoundaryValues(all grids)} \\
\>\>   \code{while (Time < ParentTime)} \\
\>\>   \code{begin} \\
\>\>\>     \code{dt = ComputeTimeStep(all grids)} \\
\>\>\>     \code{PrepareDensityField(all grids, dt)} \\
\>\>\>     \code{SolveHydroEquations(all grids, dt)} \\
\>\>\>     \code{SolveRateEquations(all grids, dt)} \\
\>\>\>     \code{SolveRadiativeCooling(all grids, dt)} \\
\>\>\>     \code{Time += dt} \\
\>\>\>     \code{SetBoundaryValues(all grids)} \\
\>\>\>     \code{EvolveLevel(level+1, Time)} \\
\>\>\>     \code{RebuildHierarchy(level+1)} \\
\>\>   \code{end} \\
\> \code{end} \\
\end{tabbing}
\end{minipage}
\caption{\ez\ AMR algorithm.}
\label{algorithm}
\end{figure}

Inside the loop which advances the grids on this level, there is a
recursive call so that all the levels with finer subgrids are
advanced as well.  The resulting order of timesteps is like the
multigrid W-cycle.

Before we update the hyperbolic gas dynamics equations and solve the
elliptic Poisson equation, we must set the boundary conditions on the
grids.  This is done by first interpolating from a grid's parent and
then copying from sibling grids, where available.  Once the boundary
values have been set, we solve the Poisson equation using the
procedure {\tt PrepareDensityField} and evolve the hydrodynamic field
equations using procedure {\tt SolveHydroEquations}. The multispecies
kinetic equations are integrated by procedure {\tt
SolveRateEquations}, followed by an update to the gas energy equation
due to radiative cooling by procedure {\tt SolveRadiativeCooling}. The
final task of the {\tt EvolveLevel} routine is to modify the grid
hierarchy to reflect the changing solution.  This is accomplished via the {\tt
RebuildHierarchy} procedure, which takes a level as an argument and
modifies the grids on that level and all finer levels.  This involves
three steps: First, a refinement test is applied to all the grids on that level
to determine which cells need to be
refined. Second, rectangular regions are chosen which cover all of the
refined regions, while attempting to minimize the number of
unnecessarily refined points. Third, the new grids are created and
their values are copied from the old grids (which are deleted) or
interpolated from parent grids. This process is repeated on the next
refined level until the grid hierarchy has been entirely rebuilt.

\subsection{Implementation}

\ez\ is written in a mixture of C++ and Fortran.
High-level functions and 
data structures are implemented in C++ and computationally intensive
lower-level functions are implemented in Fortran.  As described in more
detail below, \ez\ is parallelized
using the MPI message-passing library \cite{MPI}
and uses the HDF5 data format \cite{HDF}
to write out data and restart files in a platform-independent format.
The code is quite
portable and has been run on numerous parallel shared and distributed
memory systems, including the IBM Power N systems, SGI Altix, Cray
XT3, IBM BG/L, and numerous Beowulf-style linux clusters.

\begin{figure}
\centering
\includegraphics[height=5 cm]{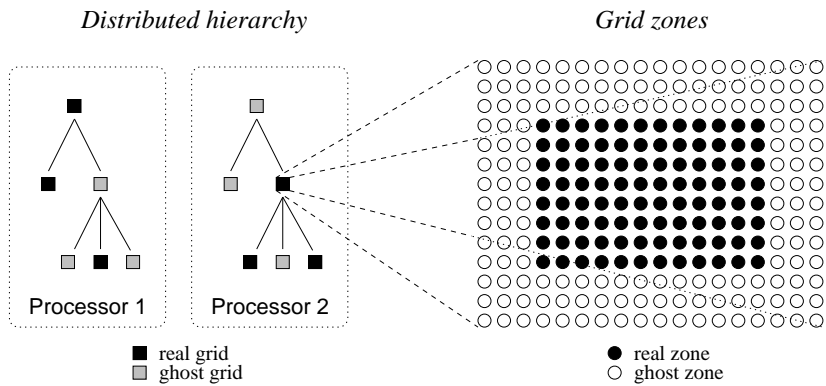}
\caption{Real and ghost grids in a hierarchy; real and ghost zones
in a grid.}
\label{fig.amrhier}
\end{figure}

The AMR grid patches are the primary data structure in \ez.  Each individual 
patch is treated as a separate object, and can contain both field variables
and particle data.  Individual patches are organized into a dynamic distributed
AMR mesh hierarchy using two different methods: the first is a hierarchical tree and
the second is a level-based array of linked lists.  This allows grids to be quickly
accessed either by level or depth in the hierarchy. The tree structure of a small illustrative 2D AMR
hierachy -- six total grids in a three level hierarchy distributed across
two processors -- is shown on the left in Figure~\ref{fig.amrhier}.


Each data field within a grid is an array of zones of 1,2 or 3 dimensions
(typically 3D in cosmological structure formation).  
Zones are partitioned into a core block of \emph{real zones}
and a surrounding layer of \emph{ghost zones}.  Real zones are used to
store the data field values, and ghost zones are used to temporarily store
neighboring grid values when required for updating real zones.
The ghost zone layer is three zones deep in order to accomodate the computational
stencil of the hydrodynamics solver, as indicated
in the right panel in Figure~\ref{fig.amrhier}.  These ghost
zones can lead to significant computational and storage overhead, especially
for the smaller grid patches that are typically found in the deeper levels of an 
AMR grid hierarchy.

\subsection{Parallelization}


Parallelization is achieved by distributing the patches, with each grid 
object locally resident in memory.  Communication between grids on 
different processors is carried out using message-passing.  
The structure of the hierarchy,
in the form of the set of linked lists described earlier, is stored redundantly 
on every processor to facilitate communication.  However, if we kept all the grid 
data on all processors that would obviously consume too much memory, 
so instead we have developed the concept of a {\it ghost grid}, which
is a nearly empty grid structure used to represent those grids that reside 
on other processors.  For every {\it real grid}, which contains all grid 
data and resides on a given processor, there are $p-1$ {\it ghost grids} 
which represent that grid on other processors (assuming p processors).  
This is feasible because 
{\it ghost grids} consume orders of magnitude less memory than 
{\it real grids} (but see below for a discussion of how this must change 
for very large numbers of processors).  This structure is shown graphically
in Figure~\ref{fig.amrhier}.

Since child grids depend on parent grids, parallelization can only 
proceed over the grids on a given level (different levels must be 
computed in a serial fashion).  Therefore, all grids on a given 
level are distributed over processors separately, starting with the 
root grid.  The root grid is simply tiled and split into a number of 
patches which is at least as large as the number of processors.  
Then, as grids are added, each grid is placed by default on the 
same processor as its parent, minimizing communication.  Once 
the rebuild of the hierarchy has completed on a given level, the 
load balancing ratio between processors is computed and grids 
are transfered between processors in an attempt to even the load.  
Because grids are discrete objects, this cannot in general be done 
in an optimal fashion (although we have experimented with grid 
splitting, see \cite{Lan2000}).  For the large problems
typical of CSF, there are many grids per processor so this is not 
typically a problem.

Communication is overlapped with computation by pre-computing 
communication pathways and starting non-blocking MPI calls.  
Care is taken to generate these as soon as possible so that the 
data will be ready when required by another processor.  This is 
one of the reasons why it is important to have the entire hierarchy 
in memory on each processor, so that all grid overlaps can be 
found and data transfer initiated early in the compute cycle.

\subsection{Fast sibling grid search}

As the code has been expanded and run on more and more 
processors with more and more grids, a number of performance 
bottlenecks have been identified and eliminated.  We describe 
one such example here.  Many binary operations between grids, 
such as copying boundary values, require first identifying which 
grids overlap.  In early versions of the code, this was done 
using a simple double-loop to perform a comparison between 
each grid and all other grids.  This is an $O(N_{\rm grid}^2)$ 
operation but because the number of operations per grid 
comparison is very small, this bottleneck did not appear until 
we ran simulations with 100's of processors, generating more 
than 10,000 grids.

The problem was solved by carrying out a chaining-mesh search 
to identify neighboring grids.  First, a coarse mesh is constructed
over the entire domain (with $4^3$ times fewer cells than the
root grid), and a linked list is begun for each chaining-mesh cell.  
Then we loop over all the grids and find which chaining-mesh 
cell(s) that grid belongs to, adding that grid to the appropriate 
linked list(s).  In this way we generate a coarse localization of 
the grids, so that when we need to find the list of neighbors for 
a given grid, we simply need to check the grids in the same 
chaining-mesh (or multiple chaining-meshes if the grid covers 
more than one)  This reduces the number of comparisons by 
a factor of about $N_{\rm chain}^{-3}$, and since the number 
of chaining mesh cells, $N_{\rm chain}$ scales with the 
problem size, the procedure reduces the operation count back 
to $O(N_{\rm grid})$ scaling.
\subsection{Enzo  I/O}

All input/output operations in \ez\ are performed using the Hierarchical Data Format 5 library (HDF5).
HDF5, specifically hdf5 v.1.6.5 has a number of very attractive features for management of large, complex
data sets and the associated metadata.  HDF5 has an excellent logical design and is available on every
major computing platform in the NSF/DOE arena.  One outstanding advantage of HDF5 is that one can 
easily re-start a model on a different computational platform without having to worry about differences
in endian-ness or internal floating point format.

The basic data model in \ez\ is that a given MPI  task ``owns" all of the I/O required by the set of top level
grids and subgrids present in that task.  A single method is used to provide checkpointing and re-start
capability and for science dumps at specified intervals.  

All I/O is designed for moving the largest possible amount of data per operation
(subject to chunking constraints described below), and all memory references are stride 1 for
maximum efficiency.  Each task performs all of its I/O independently of the other tasks and there
is no logical need for concurrency, i.e, there is actually no advantage to using the parallel
co-operative interface available in HDF5.  Although synchronization is not required in principle,
it is convenient to synchronize all processes after an entire dump has been written so that it is safe to hand
off to asynchronous processes which may move the data across a network or to archival storage.

The basic object in HDF5 is a file containing other HDF5 objects such as
data sets and associated metadata attributes.  HDF5 also supports a truly
hierarchical organization through the group concept.  \ez\ uses each of these
features.  The group concept, in particular, allows \ez\ to pack potentially
huge numbers of logically separate groups of data sets (i.e. one such set for 
each subgrid) into a single Unix file resulting in a correspondingly large
reduction in the number of individual files and making the management of
the output at the operating system level more convenient and far more 
efficient.  
In an \ez\ AMR application running on N processors with G subgrids
per MPI task with each subgrid having D individual baryon field and/or
particle lists this results in only N HDF5 files per data dump
instead of N*G files without grouping or N*G*D files in a simplistic
case with no groups or data sets and a separate file for each physical
variable. 


The packed-AMR scheme necessarily involves many seek operations
and small data transfers when the hierarchy is deep.  A vital
optimization in \ez\ is the use of in-core buffering of the assembly
of the packed-AMR HDF5 files.  This is very simply achieved using the 
HDF5 routine \texttt{H5Pset\_fapl\_core} to set the in-core buffering properties
for the file.  \ez\ uses a default buffer size of 1 MByte per file.
At present, in HDF5 v.1.6.5 this is only available on output
where increases in performance by $> 120 \times$ have been observed with {\em Lustre}
file systems.  The lack of input buffering implies that reading re-start
files can be relatively expensive compared to writing such files but
in a typical batch run there may be dozens of write operations for a
single read operation in the re-start process.  When input buffering
becomes available in a future version of HDF5 the cost of I/O operations
will be a negligible fraction of the run time even for extremely deep
hierarchies.

For very large uniform grid runs (e.g., 2048$^3$ cells and particles)
we encounter different problems.
Here individual data sets are so large it is necessary to using
chunking so that any individual read/write operation does not
exceed certain internal limits in some operating systems.
The simplest 
strategy is to ensure that access to such data sets is by 
means of HDF5 hyperslabs corresponding to a plane or several
planes of the 3D data.  For problem sizes beyond 1280$^3$ it is
necessary to use 64-bit integers to count dark matter particles
or compute top grid data volumes.  \ez\ uses 64-bit integers
throughout so it is necessary to handle MPI and HDF5 integer
arguments by explicitly casting integers back to 32-bits.

\section{Performance and scaling on terascale platforms}
\ez\ has two primary usage modes: {\em unigrid}, in which a non-adaptive uniform Cartesian grid is used, and {\em AMR}, in which adaptive
mesh refinement is enabled in part or all of the computational volume. In actuality,
unigrid mode is obtained by setting the maximum level of refinement to zero in an
AMR simulation, and precomputing the relationships between the root grid tiles. Otherwise, both calculations using the same machinery in \ez\ and are updated according to Fig. \ref{algorithm}. The scaling and performance of \ez\ is very different in these two modes, as are the memory and communication behaviors. Therefore we present both in this section. 

\subsection{Unigrid application}

\begin{figure}[performance]
\begin{center}
\subfigure[\label{UGa}]{\includegraphics[angle=-90,width=9cm]{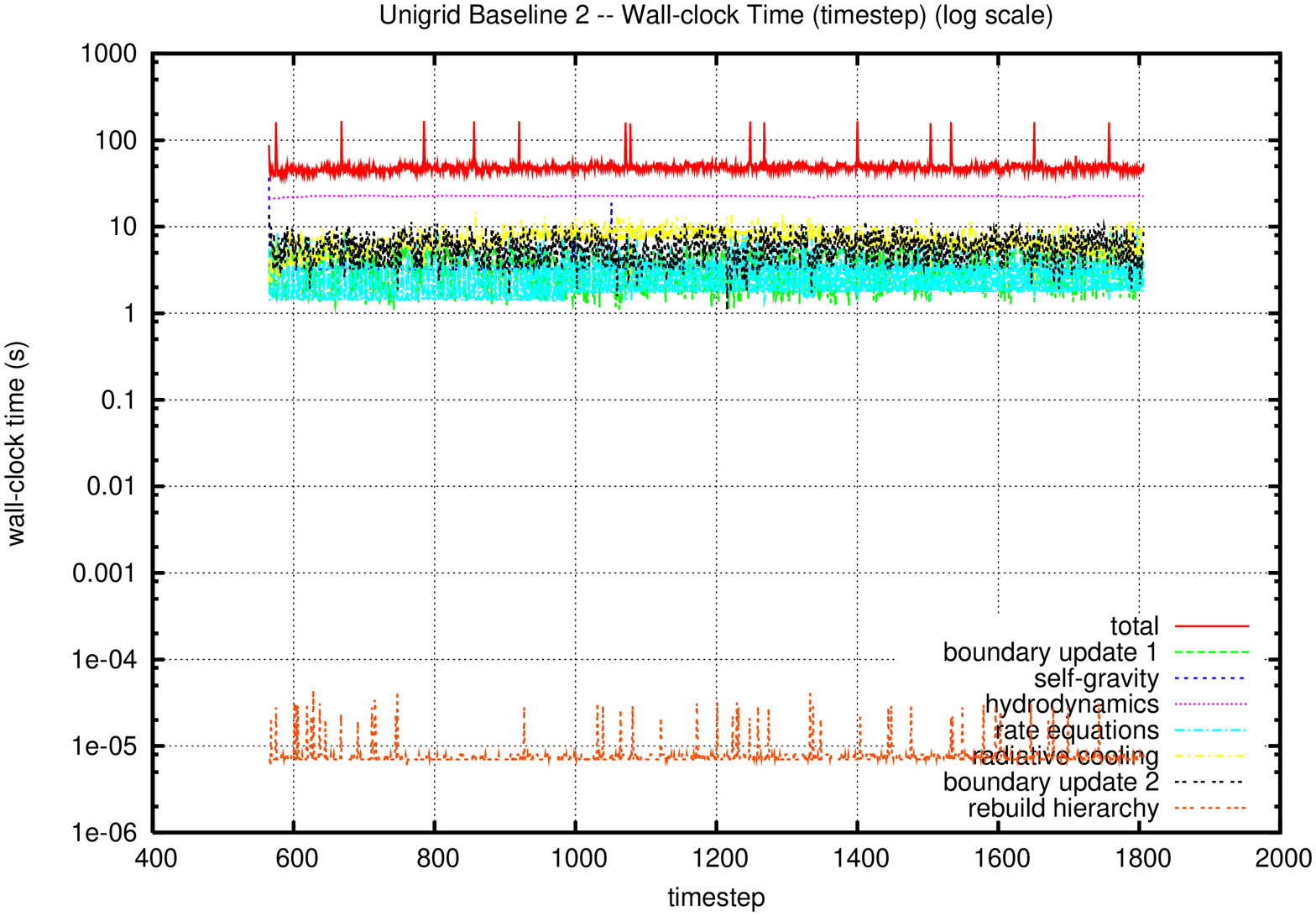}}
\subfigure[\label{UGb}]{\includegraphics[angle=-90,width=9cm]{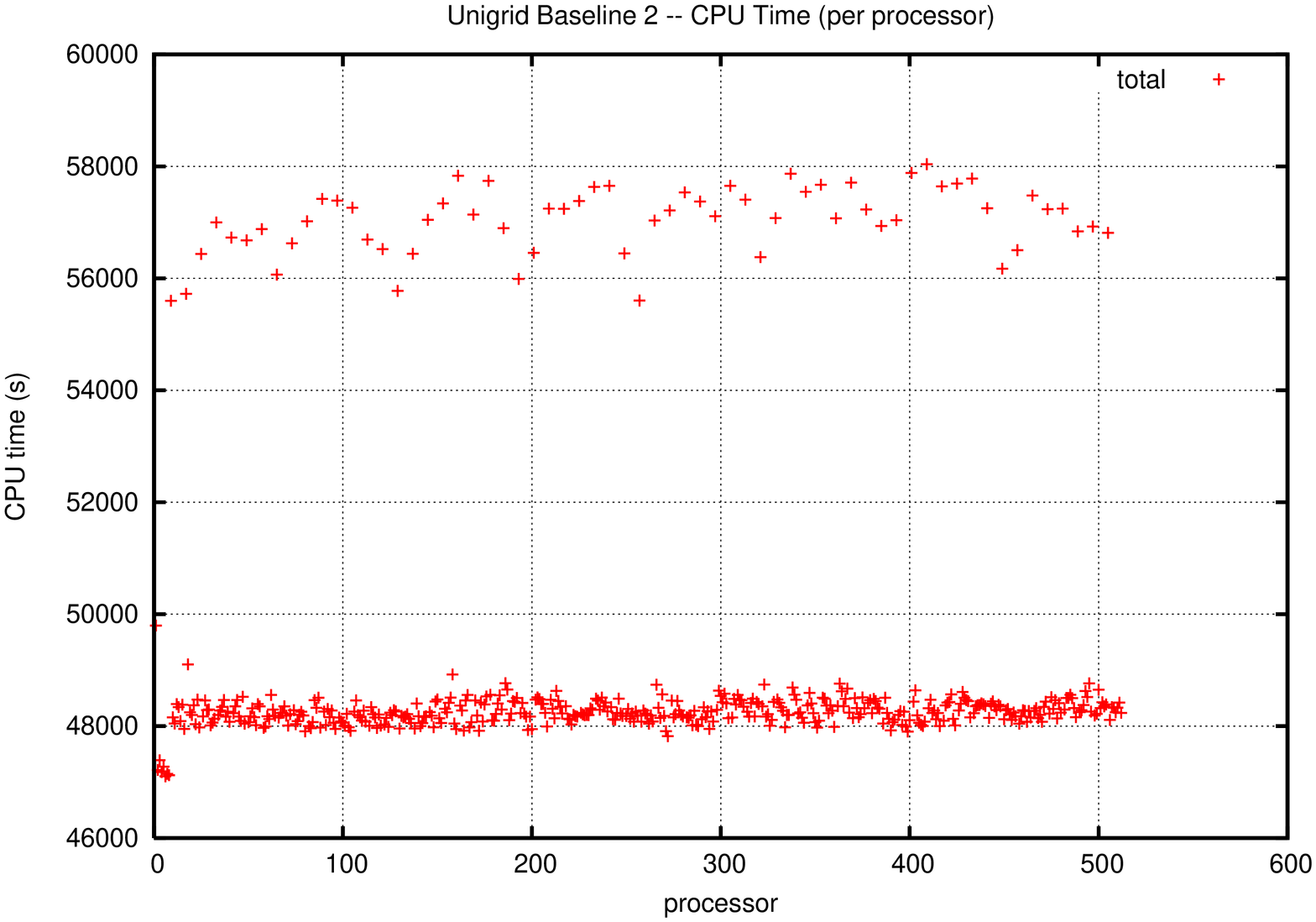}}
\end{center}
\caption{(a) Wall-clock time per timestep versus timestep broken down according to 
major code region for the $1024^3$ unigrid simulation. Spikes in total time correspond
to I/O events.; (b) CPU time per processor versus processor. Simulations were carried out on 512 cpus on NERSC IBM Power 5 system {\em Bassi}.}
\end{figure}

Fig. \ref{UGa} shows the results of a recent unigrid simulation carried out 
on 512 processors of NERSC {\em bassi}, and IBM Power5 system. The simulation was 
performed on a grid of $1024^3$ cells and the same number of dark matter particles.
The simulation tracked 6 ionization states of hydrogen and helium including
nonequilibrium photoionization and radiative cooling. Fig. \ref{UGa} plots the total wall-clock time per timestep versus timestep as well as the cost of major code regions
corresponding to the procedure calls in Fig. \ref{algorithm}. Unigrid performance is quite predictable, with the cost per timestep for the different code regions being roughly constant. In this example hydrodynamics dominates the cost of a timestep (45\%). The remaining cost breaks down as follows: radiative cooling (15\%),  
boundary update 2 (11\%), self-gravity (7\%), boundary update 1 (7\%), rate equations (3\%). An additional 12\% is spent in message passing. Fig. \ref{UGb} shows the 
CPU time per processor for the job. We find that the master processor in each 8 processor
SMP node is doing 9\% more work than the other seven, otherwise the workload is nearly
uniform across processors.

\begin{figure}[performance]
\begin{center}
\subfigure[\label{AMRa}]{\includegraphics[angle=-90,width=9cm]{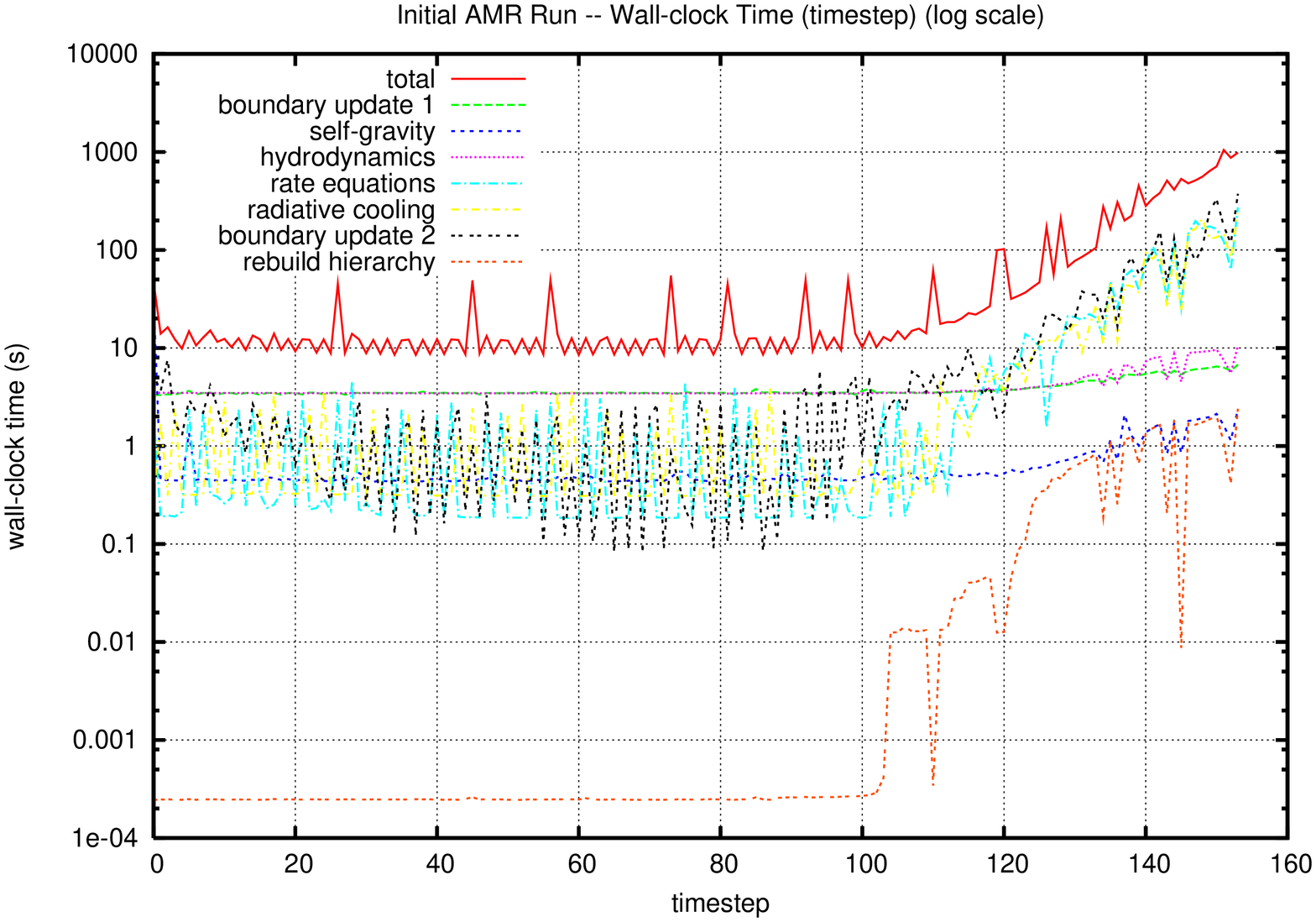}}
\subfigure[\label{AMRb}]{\includegraphics[angle=-90,width=9cm]{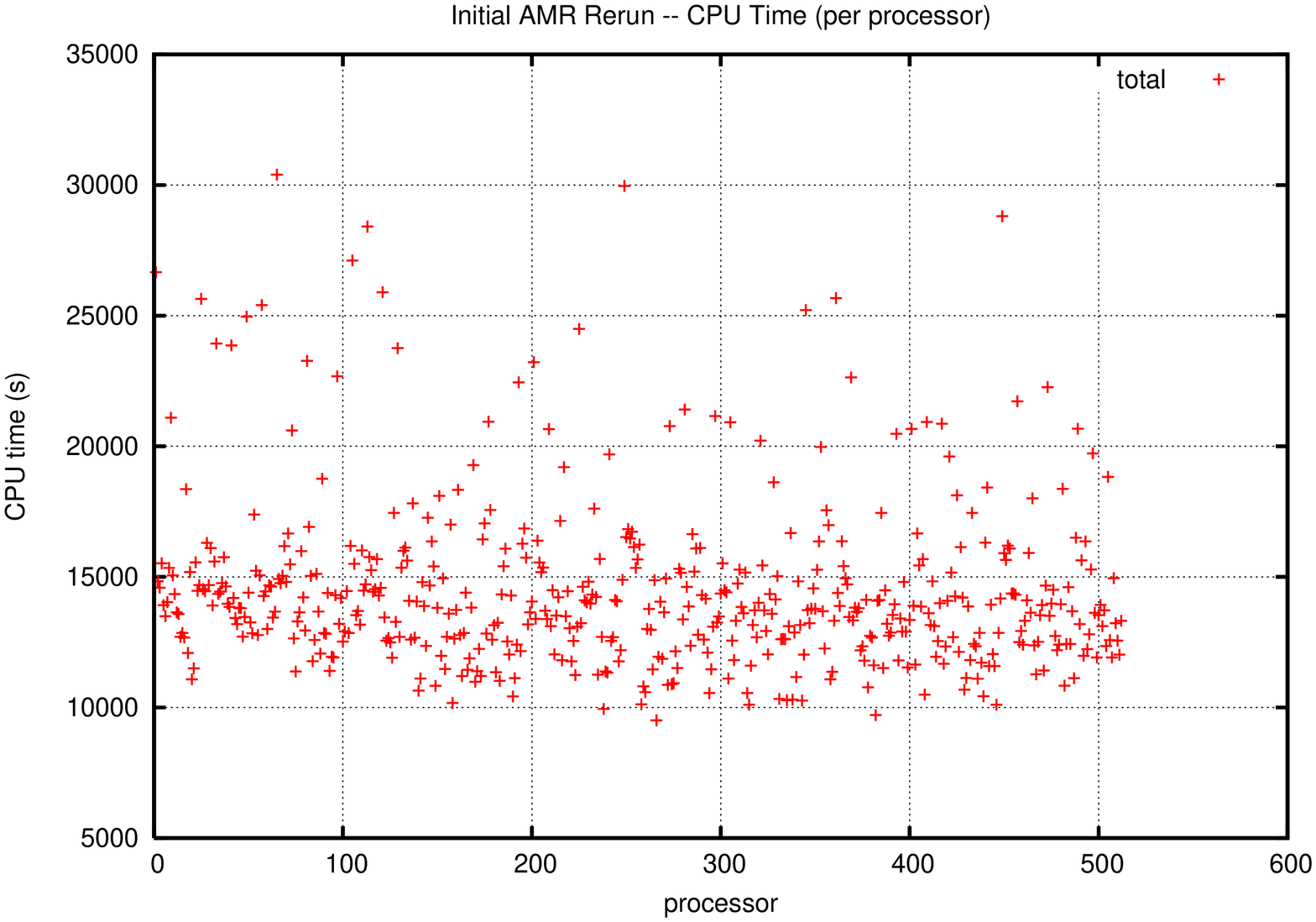}}
\end{center}
\caption{(a) Wall-clock time per root grid timestep versus root grid timestep broken 
down according to 
major code region for the $512^3$ 4-level AMR simulation; (b) CPU time per processor versus processor. Simulations were carried out on 512 cpus on NERSC IBM Power 5 system {\em Bassi}.}
\end{figure}

\subsection{AMR application}

Fig. \ref{AMRa} plots the wall-clock time per root grid timestep versus timestep for 
an AMR simulation with identical cosmology and physics as our unigrid example, 
only here the base
grid has dimensions $512^3$ and up to 4 levels of refinement are permitted in
dense regions. Prior to the onset of mesh refinement at timestep 103
the AMR simulation behaves 
identically to a unigrid simulation, with hydrodynamics dominating the cost
per timestep. By timestep 156, the cost of a root grid timestep has increased
one hundred-fold. There are several reasons for this. First, we
use hierarchical timestepping in \ez, which means that for each root grid timestep, a subgrid of level $\ell$ will be timestepped roughly $2^{\ell}$ times. The total number
of substeps per root grid timestep for a fully refined region
is $\sum_{\ell=1}^{\ell_{max}} 2^{\ell}$, which is
30 for $\ell_{max}$=4. By timestep 156, only a small fraction
of the total volume is refined, and the average number of substeps per root grid
timestep is 8, far less than the factor 100 we observe. The dominant cause for the
upturn is the cost of procedure {\tt boundary\_update\_2}, which exchanges ghost zone
data between every subgrid at every level and substep. Secondary causes for
the upturn are the procedures
{\tt rate\_equations} and {\tt radiative\_cooling}. 
These routines are both subcycled on a chemical
timescale that is short compared to the hydrodynamic time step. The separation
between these two timescales increases as the gas density increases. AMR allows the
gas to become very dense, and thus the ionization of cooling calculations grows
to dominate the hydrodynamics calculation. 

Fig. \ref{AMRb} shows the
cumulative CPU time as a function of processor number. There is considerably
more spread between the largest and smallest time (30000/10000=3) compared with the unigrid run (5800/4800=1.21). This is due to true load imbalances arising from
mesh refinement, chemistry/cooling subcycling, and communication loads. At present our dynamic load balancing algorithm in \ez\ only trys to equalize the zone-timesteps among processors. It does not attempt to balance communications and subcycling loads. This is
an obvious area for improvement. 

\subsection{Parallel scaling}
Because the cost per timestep of a unigrid simulation is roughly constant,
varying by only a factor of a few as physical conditions change, parallel
scaling tests are easy to perform for even the largest grids and particle
counts. The same is not true of AMR simulations since the load is constantly
changing in a problem-dependent way. Time--to--solution is the only robust metric.
Because large AMR simulations are so costly, it is not possible to run many cases to find the optimal machine configuration for the job. Consequently we do not have parallel
scaling results for AMR simulations of the size which require terascale 
platforms (for smaller scaling tests, see \cite{Bordner07}.) In practice, processor counts are driven by memory considerations,
not performance. We expect this situation to change in the petascale era where
memory and processors will be in abundance. Here, we present our unigrid parallel scaling results for \ez\ running on a variety of NSF terascale systems available to us.

\begin{figure}
\centering
\includegraphics[width=8 cm]{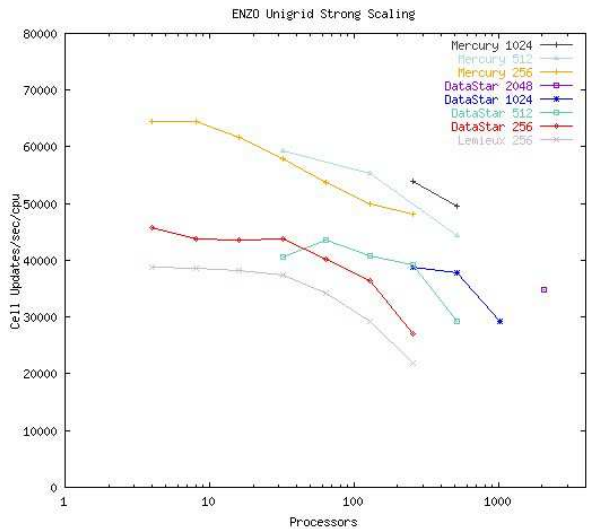}
\caption{\ez\ strong scaling results for unigrid cosmology runs of size
$256^3, 512^3, 1024^3$ and $2048^3$ on NSF terascale platforms.}
\label{scaling}
\end{figure}

Figure \ref{scaling} shows the results of strong scaling tests of \ez\
unigrid simulations of the type described above for grids of size $256^3, 512^3, 1024^3$ and $2048^3$ and an equal number of dark matter particles. We plot cell updates/sec/cpu
versus processor count for the following machines: {\em Lemieux}, a Compaq DEC alpha cluster at PSC, {\em Mercury}, an IBM Itanium2 cluster at NCSA, and
{\em DataStar}, an IBM Power4 cluster at SDSC. Ideal parallel scaling
would be horizontal lines for a given archictecture, differentiated only by their 
single processor speeds. We see near-ideal scaling for the $256^3$ test on
{\em DataStar} and {\em Lemieux} up to 32 processors, followed by a gradual roll-over in parallel performance to $\sim 50\%$ at 256 processors. Nonideality sets in at 16 processors on {\em Mercury}, presumably due to its slower communications fabric. 
As we increase the problem size on any architecture, parallel efficiency
increases at fixed NP, and the roll-over moves to higher NP. Empirically, we find
that parallel efficiency suffers if grid blocks assigned to individual
processors are smaller than about $64^3$ cells. Using blocks of size $128^2 \times 256$, we find that the cell update rate for
our $2048^3$ simulation on 2048 {\em DataStar} processors is $\sim 80 \%$ the cell
update rate of our $256^3$ on 4 processors. \ez\ in unigrid mode is very scalable.

\section{Toward petascale Enzo}

\subsection{New AMR data structures}

\newcommand{\numgrids}{n}
\newcommand{\numprocs}{p}
\newcommand{\sizegrid}{|G|}
\newcommand{\sizegridlocal}{|G_l|}
\newcommand{\sizegridremote}{|G_r|}
\newcommand{\sizefield}{|F|}
\newcommand{\sizefields}{\sum_i^n |F_i|}

For moderate numbers of processors, the current datastructure used for
storing the AMR grid hierarchy is adequate.  Even though the hierarchy
topology is stored redundantly on each processor, because the data
fields are vastly larger than \ez's individual C++ \code{grid}
objects, the extra memory overhead involved is insignificant.
However, as the number of processors increases, this memory overhead
increases as well.  For the processor counts required for
petascale-level computing, the storage overhead would overwhelmingly
dominate, to the point where the largest computation would actually
not be limited by the number of processors, but rather by the amount
of memory available for each processor.  Thus, for \ez\ to scale to
the petascale level, the memory overhead for storing the AMR hierarchy
must be reduced.

The memory required for storing \ez's current AMR
datastructure can be approximated as $\sizefield + \numgrids \numprocs
\sizegrid$, where the first term $\sizefield$ is the field
variable data, and the second term $\numgrids \numprocs \sizegrid$ is
the overhead for storing the grid hierarchy datastructure.  Here
$\numgrids$ is the number of grid patches, $\numprocs$
is the number of processors, and $\sizegrid$ is the storage required
by a C++ \code{grid} object.

One approach to reducing the size of the overhead term $\numgrids \numprocs
\sizegrid$ is to reduce the size of the \code{grid} object by removing
unnecessary member variables from the \code{grid} class.  Since some
member variables are constant for a hierarchy level, and some are
constant for the entire hierarchy, the amount of memory required would
be reduced.  This refactoring is indicated by the first transition in
Figure~\ref{f:enzodata1}.  Although this would indeed save some
memory, preliminary estimates indicate that the savings would only be
about $13.5\%$.

\newlength{\figwidth}
\newlength{\figheight}
\newlength{\figindent}
\setlength{\figwidth} {0.200\textwidth}
\setlength{\figheight}{0.1375\textheight}
\setlength{\figindent}{0.65\figwidth}
\begin{figure}
\begin{minipage}{\textwidth}
\begin{center}
\begin{minipage}{\figwidth}
\epsfig{file=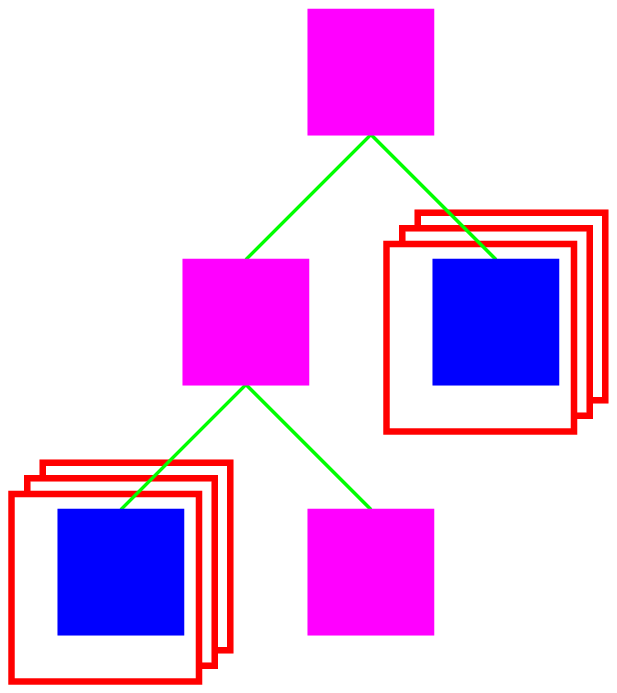,width=\figwidth,height=\figheight}
\end{minipage}
\begin{minipage}{0.25in}
\centerline{$\Rightarrow$}
\end{minipage}
\begin{minipage}{\figwidth}
\epsfig{file=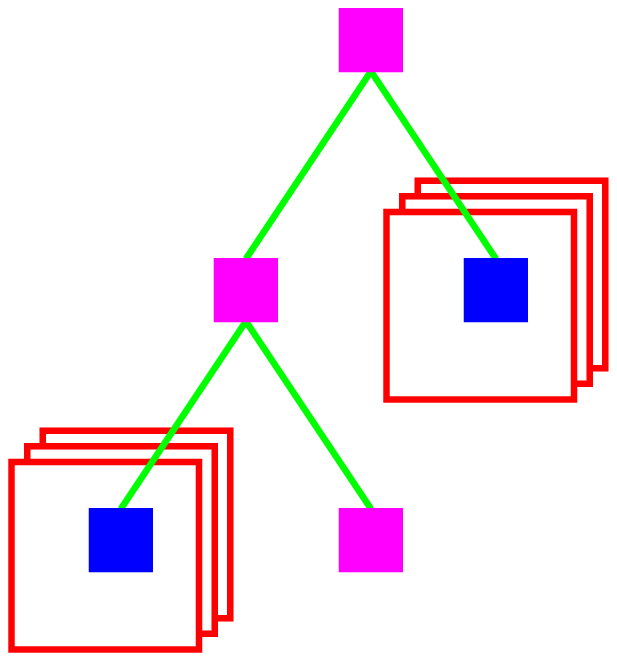,width=\figwidth,height=\figheight}
\end{minipage}
\begin{minipage}{0.25in}
\centerline{$\Rightarrow$}
\end{minipage}
\begin{minipage}{\figwidth}
\epsfig{file=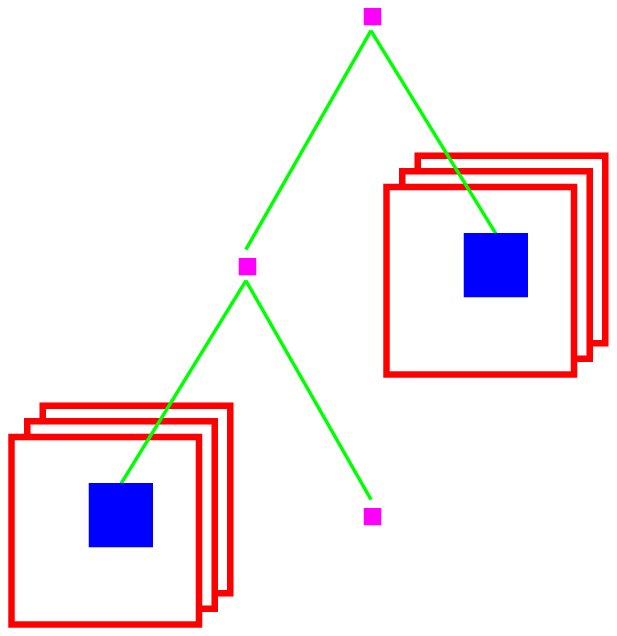,width=\figwidth,height=\figheight}
\end{minipage}
\end{center}
\end{minipage}
\caption[Reducing AMR datastructure storage overhead]
        {Reducing AMR datastructure storage overhead.  Solid blue 
        squares represent local \code{grid} objects, solid magenta 
        squares represent remote \code{grid} objects, and open red boxes
        represent data fields.  The first transition illustrates reducing the
        size of all \code{grid} objects, and the second transitions illustrates
        splitting the single \code{grid} class into two \code{grid\_local} and \code{grid\_remote} subclasses.}
\label{f:enzodata1}
\end{figure}

A second approach to reducing the size of the overhead term would be
to split the \code{grid} class into two subclasses \code{grid\_local}
and
\code{grid\_remote}, and use  \code{grid\_local} objects for local grids
that contain field data, and \code{grid\_remote} objects for grid
patches whose data fields reside on another processor.  This
refactoring is indicated by the second transition in
Figure~\ref{f:enzodata1}.  This modification would change the overhead storage term from
$\numgrids
\numprocs \sizegrid$ to $\numgrids ((\numprocs-1)
\sizegridremote + \sizegridlocal$), where $\sizegridremote$ is the
size of the \code{grid\_remote} class and $\sizegridlocal$ is the size
of the \code{grid\_local} class.  The advantage of doing this is that the 
\code{grid\_remote} class could be made much smaller, since most of
the variables in the \code{grid} class are only required for local
grids.  Also, a vast majority of the grid classes are these much
smaller \code{grid\_remote} objects.  Thus the memory savings for this
modification would be quite large---a factor of roughly $14$ over the
already slightly reduced \code{grid} size from the first approach.

\subsection{Hybrid parallelism}

Another improvement would of course be not to store the entire grid
hierarchy on each processor.  The easiest approach would be to store
one copy per shared memory node, which would decrease the memory
storage further by a factor equal to the number of processors per
node. This could be implemented with a hybrid parallel programming
model in which instead of assigning one MPI task per processor
which serially executes its root grid tile and all its subgrids,
we assign one MPI task per node. This heavy weight node would
execute a somewhat larger root grid tile and its more numerous subgrids,
using shared memory parallelism wherever possible. For example,
every subgrid at a given level of refinement would be processed
concurrently using OpenMP threads \cite{OMP}.

While the above modifications to the AMR datastructure should allow
\ez\ to run on machines with on the order of $10^4$ to $10^5$
processors, extending to $10^6$ processors would require reducing the
overhead even further.  The ultimate improvement memory-wise would be
to store a single copy of the grid hierarchy, though depending on the
node interconnect that would cause a communication bottleneck.  A
refinement on this approach would be to store one copy of the
hierarchy for every $M$ processors, where $M$ is some
machine-dependent number chosen to balance the tradeoff between memory
and communication overhead.

\subsection{Implicitly-coupled radiation hydrodynamics}

We are working to incorporate radiation transfer processes within the
\ez\ framework to improve the physical realism of cosmological
modeling of self-regulated star formation and predictions on the epoch
of cosmic re-ionization.  These efforts are unique within the
computational astrophysics community, because unlike traditional
approaches to such multi-physics couplings, we are coupling the
radiation transfer implicitly with both the fluid energy and
chemical kinetics processes.  This approach promises to provide a
fully-consistent coupling between the physical processes, while
enabling the use of highly scalable solvers for the coupled solution.

Through implicitly coupling the radiation-hydrodynamics-chemical
kinetics processes together, we have the benefits of numerical
stability in time (regardless of step size) and the ability to use
high-order time discretization methods.  On the other hand, this
coupling results in a nonlinear system of partial differential
equations that must be solved at each time step.  For this coupled
solution we will use {\em Inexact Newton Methods}, which in recent
years have been shown to provide an efficient and scalable approach to
solving very large systems of nonlinear equations
\cite{KnollKeyes2004,KeyesReynoldsWoodward2006}.  This scalability
arises due to a number of factors, notably the fact that for many
problems Newton's method exhibits a convergence rate that is
independent of spatial resolution, so long as the inner linear solver
scales appropriately \cite{WeiserEtAl2005}.  As radiation diffusion
processes are elliptic in nature, and radiative couplings to fluid
energy and chemical kinetics occur only locally, we plan to achieve
such optimal scalability in the linear solver through the use of a
Schur Complement formulation (\cite{BrownWoodward2001}) to reduce the
coupled Newton systems to scalar diffusion problems, which will then
be solved through optimally scalable multi-level methods, provided by
the state-of-the-art HYPRE linear solver package
\cite{FalgoutYang2002,hypre_site}.  This solver library has been shown
to scale up to massively parallel architectures
(\cite{BakerFalgoutYang2006}), and as the computational heart of the
Inexact Newton approach is the inner linear solver, we are confident
that such an implicit formulation and solution will enable
radiation-hydrodynamics-chemical kinetics simulations in \ez\ to the
Petascale and beyond. 

\subsection{Inline analysis tools}

Petascale cosmology simulations will provide significant data analysis challenges, primarily due to the size of simulation datasets.  For example, an \ez\ Lyman alpha forest calculation on a $1024^3$ grid, the current state-of-the-art, requires approximately 110 GB of disk space per simulation output.  Tens or hundreds of these outputs are required per simulation.  If scaled to a $8192^3$ grid, a reasonable size for a petascale computation, this would result in 56 TB of data per simulation output, and multiple petabytes of data written to disk in total.  Given these dataset sizes, analysis can become extremely time-consuming: doing even the simplest post facto analysis may require reading petabytes of data off of disk.  For more complex analysis, such as spectrum or N-point correlation function generation, the data analysis may be comparable in computational cost to the simulation itself.  Analysis tools for petascale datasets will by necessity be massively parallel, and new forms of data analysis that will be performed hundreds or thousands of times during the course of a calculation may make it exceedingly cumbersome to store all of the data required.  Furthermore, it will become extremely useful to be able to monitor the status and progress of petascale calculations as they are in progress, in an analogous way to more conventional experiments.

For these reasons, among others, doing data analysis while the simulation itself is running will become necessary for petascale-level calculations.  This will allow the user to save disk space and time doing I/O, greatly enhance the speed with which simulations are analyzed (and hence improve simulation throughput), and have access to data analysis strategies which are otherwise impractical or impossible.  Analysis that can be done during cosmological simulations include the calculation of structure functions and other global snapshots of gas properties; dark matter power spectra, halo and substructure finding, and the generation of merger trees; radial profile generation of baryon properties in cosmological halos; production of projections, slices, and volume-rendered data for the creation of movies; the generation of synthetic observations, or ``light cones"; the calculation of galaxy population information such as spectral energy distributions and color-magnitude relations; global and halo-specific star formation rates and population statistics; ray tracing for strong and weak lensing calculations and for Lyman alpha forest or DLA spectrum generation; and essentially any other type of analysis that can be massively parallelized.  Analysis of simulations in this way, particularly if done with greater frequency in simulation time, may enhance serendipitous discovery of new transient phenomena.

A major constraint to inline analysis of cosmological simulations is that it requires very careful and possibly time-consuming planning on the part of the scientists designing the calculation, and is limited to analysis techniques that can be heavily parallelized.  As a result, inline analysis techniques will not completely negate the necessity of writing out significant amounts of data for follow-up analysis and data exploration.  Furthermore, inline analysis will require careful integration of analysis tools with the simulation code itself.  For example, the simulation and analysis machinery will, for the sake of efficiency, require shared code, data structures, and parallelization strategies (i.e. domain-decomposed analysis if the simulations are similarly decomposed).  If these relatively minors hurdles can be surmounted, however, this sort of analysis will result in gains far beyond the additional computational power used.

\section{Acknowlegements}
\ez\ has been developed with the support of the National Science Foundation via grants ASC-9318185, AST-9803137, AST-0307690, and AST-0507717; the National Center for Supercomputing Applications via PACI Subaward ACI-9619019; and the San Diego Supercomputer Center via the Strategic Applications Partners program. Greg Bryan acknowledges support from NSF grants AST-05-07161, AST-05-47823 and AST-06-06959.

\end{document}